# A review of experiments reporting non-conventional phenomena in nuclear matter aiming at identifying common features in view of possible interpretation

Stefano Bellucci[1], Fabio Cardone[2,3] and Fabio Pistella[4]

[1]*INFN-Laboratori Nazionali di Frascati, Via E. Fermi 54, 00044 Frascati, Italy*
[2]*GNFM, Istituto Nazionale di Alta Matematica "F.Severi"*
*Città Universitaria, P.le A.Moro 2 - 00185 Roma, Italy*
[3]*Istituto per lo Studio dei Materiali Nanostrutturati (ISMN – CNR)*
*c/o Università La Sapienza di Roma - 00185 Roma, Italy*
[4]*Ente Nazionale di Ricerca e Promozione per la Standardizzazione (ENR), Italy*

### ABSTRACT

The purpose of the present paper is to clarify, as far as it is possible, the overall picture of experimental results in the field of non-conventional phenomena in nuclear matter published in scientific literature, accumulated in the last decades and still missing a widely accepted interpretation. While completeness of the collection of the experiments is not among the aims of the effort, focus is put on adopting a more comprehensive and integral approach through the analysis of the different experimental layouts and the different results, searching for common features and analogous factual outcomes in order to obtain a consistent reading of a lot of experimental evidences that appear, until now, lacking a classification in a logic catalogue which might be compared to a sort of building and not to a collection of single stones. Particular attention is put on the issue of reproducibility of experiments and on the reasons why such a limitation is a frequent characteristic of many experimental activities reported in published papers. This approach is innovative as compared with those already available in the scientific literature

In a synoptical table a comprehensive classification is given of the twenty experiments examined in terms of type of evidences that are ascertained by the experimenters in their published papers but are "unexpected" according to well established physical theories. Examples of such unexpected evidences (named also non-conventional or weird) are: excess heat generation, isotope production, reduction of radioactivity levels, and production of neutrons or alpha particles. These evidences are classified taking into account both the material where the evidence takes place (Solutions, Metals, Rocks and Artificial materials) and the stimulation tecniques (supply of electric voltage, irradiation by photons, mechanical pressure) used to generate the evidences (which do not appear in absence of such stimuli at an appropriate intensity).

As an Appendix, "identity cards" are provided for each experiment examined, including details emerged during the experiment and reported in each pertaining paper, that sometimes are not given adequate consideration neither by the author of the experiment nor in other review papers. The analysis of the details provides suggestions (also referred to as clues in this papers) used to formulate the content of the second part of each identity card where inferences deduced from facts are outlined in view of presenting tentative interpretation at microscopic level. This is done by concentrating attention on the clues repeated in different experiments in order to yield possible explanations of the "unexpected" evidences.

The main outcome of such analysis is that, in all examined cases, a common "operation" can be identified: the stimulation techniques mentioned above can be interpreted as a sort of compression[1] producing a ramp of energy densification (with reference to volumes in space, or time coordinates). Five types of densifications were identified. This reading in terms of energy densification is in accordance with the previsions of the Deformed Space Time theory, reported in scientific literature, in the contest of a generalization of the Einstein relativity theory, according to which the existence of energy thresholds is found to separate, for each interaction, the flat metric part from the deformed metric part and the appearance of new microscopic effects as a consequence of trespassing such thresholds. The phenomena occurring in the deformed part of the interaction metric are governed by the energy

---

[1] We use the term "compression" to indicate the operation activated by the experimenter; as such it is objective. We consider energy densification an inference of possible consequences of the operation on the status of the system

density in the spacetime (volume and time interval). This energy density is computed from the threshold energies and is peculiar to the phenomenology under consideration.

As a conclusion it is suggested that the qualified information dug out, homogenized and elaborated on, might help in repeating with proper adjustments and adequate additional instrumentation, some key experiments, in order to ensure systematic reproducibility, which is a prerequisite for interpretations and explanations to be sound and credible, as well in deriving from such an effort, indications for new experiments. It is not comfortable that, after thirty years, questions are still pending to which the most acknowledged physical theories are not capable of giving an answer. Even a definitive demonstration that all these experiments have decisive faults would be preferable than leaving the issue unaccounted for. Major research Agencies, for instance in USA and in Europe are moving in this direction.

# 1. INTRODUCTION

## 1.1 LENR: AN OPEN CONTROVERSY IS UNDER SCRUTINY AGAIN

### 1.1.1 Early phase

It is well-known that the issue of Low-Energy Nuclear Reactions (LENR)[2], has a complex, controversial history, since the original work in 1989 [1] which generated diffused expectations for the feasibility, by taking advantage of previously unknown nuclear phenomena, of energy production at low cost and without negative side effects. The first phase was characterized by a huge amount of attempts in the scientific community and in industrial companies but It led to a disappointing early dismissal by the majority of the scientific research community; a typical example of criticism is the paper by Huizenga published in 1993 [2].

### 1.1.2 Three decades of experimental efforts

The LENR experiments conducted later, an effort which nevertheless unavoidably continued for decades in various parts of the world - a list of papers can be found in Ref. [3] - never were recognized to have achieved the high level of rigor and repeatability, which characterizes accepted modern science.

Among the most cited reviews are those published in the years 2007 to 2016 by two authors: Edmund Storms [4], [5] and Steven B. Krivit [6], [7], [8] both very skeptic about the quality of the results obtained by the experimenters engaged in this field. Nevertheless, activities on this subject continued.

Three updated and extended reviews have been recently published. In a paper published on Nature in 2019 [9] C. P. Berlinguette and others recognize that their "efforts, have yet to yield any evidence of cold fusion". Nevertheless, they "believe that there is exciting new science to be done within the parameter space of cold fusion experiments, and that this is an area worthy of engagement from the broader scientific community, even if the discovery of cold fusion at high enough rates for energy applications does not materialize. In the review paper by Nagel [10] besides references of several experimental papers, some classification of types of experiments is also given. The main limitation is that only experiments involving deuterium or hydrogen interacting with palladium or nickel are concerned. The most recent review (December 2021) is in the paper by L. O. Freire and D. Andres de Andrade [11] who, after listing a large share of the experiments published in literature, avoided to formulate any sort of conclusion and closed their paper with two entirely open questions: the first "whether the golden dream of the fusion is approaching our time; the second, whether this is a dream or a nightmare. In case cold fusion becomes a reality, perhaps the answer to the second depends on the time that technology development starts".

---

[2] Since 1989 for three decades, the anomalous transformations of matter have gone under multiple names that have increased confusion concerning phenomena that are already complex to understand; we mention a few frequently used names: Cold fusion, Low-Energy Nuclear Reaction (LENR), Condensed Matter Nuclear Physics, Nuclear metamorphosis. Piezonuclear reactions was also used as a consequence of experiments where unexpected evidences occurred when pressure increase was involved. We chose the last one which seemed to us less equivocal and more responsive to the need to give a genuinely new name to a genuinely new phenomenon. Indeed, a series of experimental results on transformation of matter which are well documented in scientific literature give rise to undeniable contradictions with established physical theories and have not been interpreted yet through models that have reached a widespread consensus.

Of some interest is also a list of papers named *Synopsis of Refereed Publications on Condensed Matter Nuclear Reactions* (2019) [12] which is more a bibliographical list than a synopsis since for each paper only journal, title, authors, and affiliations are reported, with very limited indications of the results obtained and no detailed discussion. A site named LENR-CANR [3] (already mentioned) features a library of papers on LENR including more than 1,900 original scientific papers. The papers are linked to a bibliography of over 4,500 journal papers, news articles and books about LENR.

It's worth mentioning initiatives dealing with LENR undertaken by official research Agencies at national and international level.

The most engaging effort is the one conducted under the auspices of the European Commission [13] through a Program named *Clean Energy from Hydrogen-Metal Systems* [14] responding to the call for proposals dealing with the objective *Developing a new source of clean energy*. The program started in 2020 with a duration of 4 years and a budget of 5.5 million euros. A possible extension due to the delays imposed by the COVID pandemic is under discussion. In execution of this program a large number of papers have been published and are being published in scientific journals including Journal of Condensed Matter Nuclear Science [15], specialized on this theme, to which an international Conference is dedicated annually since more than twenty years [16]. Discussion on the possible prosecution of this effort are still pending.

As an example of attention paid also recently on the issue of LENR within qualified scientific institutions, one might mention a paper [17], published by NASA in 2020, dealing with "novel" nuclear reactions observed in bremsstrahlung-irradiated deuterated metals.

Attention to the issue of LENR has been repetitively paid by US institutions in the realm of defense research activities under the responsibility of DARPA. A hearing on the subject of LENR, was held in 2016 on request of the *House Committee on Armed Services* (US Congress) organized by the Department of Defense. The outcomes of the hearing, reported in [18], focused on two methods considered as more extensively addressed by researchers: muon catalytic fusion and electrolysis. As a consequence of the political discussion the foundation of ARPA-E was established.

This effort has not led until know to conclusions on the issue of LENR: from the experimental viewpoint unpredicted events of different type continue to be observed, but reproducibility is poor and a convincing explanation of the phenomena is missing even at a phenomenological level.

### 1.1.3 Proposals formulated for theoretical interpretations

Before dealing with further developments on LENR it's worth wile to formulate a few remarks on the situation of the theoretical considerations on the issue of LENR.

A large number of theoretical models for LENR are commented by V. A. Chechin, et alii in their paper dated 1994 entitled *Critical review of theoretical models for anomalous effects in deuterated metals* [19]. The major limit of this review is that only interpretations dealing with deuterium are taken into consideration. A more recent (started in 2012) collection of papers dealing with Low Energy Nuclear Reactions can be found in the site *New Energy Times* under Low-Energy Nuclear Reactions (LENR) Theory Index [20] where links are given to a variety of specific papers and mention is made to a categorization of the theories in several classes, suggested by V. A. Kirkinskii, Yu. A. Novikov "Theoretical modelling of Cold Fusion", Novosibirsk 2002, cited as Reference 9 in Reference [21].

Among the approaches suggested for theoretical interpretation of the experimental results mention should be made of a theoretical model [22] (named MFP after the proponents Mignani, Francaviglia, Pessa) based on the Deformed Minkowski Space also named Deformed Spacetime – DST. This model has been applied by us in a recent paper [23] (published in *Symmetry*) comparing its predictions with a limited set of experimental data available in literature that were considered to be more accurately reported in the corresponding papers: the predictions have resulted to be coherent with measured outcomes.

Even if it is undeniable that the issue of LENR is continuing to be given widespread attention, it must be recognized that the picture emerging from the review engagement deployed until now is far from being clear at phenomenological level (even reproducibility is lacking in many cases) and that attempts to deepen interpretation of experimental evidences and to indicate a theoretical explanation are very diversified and even controversial. Such a situation is to be called an open question that should not be left as such.

## 1.2 A revival of attention in view of a definitive clarification

### 1.2.1 A new strategic approach adopted in USA

Recently ARPA-E addressed again the issue of LENR after the non-enthusiastic conclusion of the Hearing at US Congress level held in 2016. A workshop was organized in 2021 [24] with the purpose to "explore compelling R&D opportunities in Low-Energy Nuclear Reactions (LENR), in support of developing metrics for a potential ARPA-E R&D program in LENR. Despite a large body of empirical evidence for LENR that has been reported internationally over the past 30+ years in both published and unpublished materials, as well as multiple books, there still does not exist a widely accepted, on-demand, repeatable LENR experiment nor a sound theoretical basis. This has led to a stalemate where adequate funding is not accessible to establish irrefutable evidence and understanding of LENR, and lack of the latter precludes the field from accessing adequate funding."

Subsequently ARPA-E asked [25] "outstanding scientists and engineers from different organization, scientific disciplines, and technology sectors to participate in an Exploratory Topic (dealing with LENR both on scientific aspects and on implementation capabilities). Multidisciplinary and cross-sector collaboration spanning organizational boundaries enables and accelerates the achievement of scientific and technological outcomes that were previously viewed as extremely difficult, if not impossible. On September 2022 a call for proposal [26] was issued by ARPA-E and on February 2023 the selection of 8 project teams was decided for a total funding of 10 million dollars [27]. It's interesting to read the synthetic motivation for this effort published by ARPA: "The teams announced today are set out to answer the question 'does this area show promise, and if so, how? Or can we conclusively show that it does not?' While others have shied away from this space, ARPA-E wants to break through the knowledge impasse and deepen our understanding."

A commented chronicle of the steps undertaken by ARPA-E has been published by ANS (American Nuclear Society) [28] under the title "ARPA-E picks eight teams to prove - or debunk - low-energy nuclear reactions".

### 1.2.2 The contribution by the present paper: a collection of experimental data organized in a coherent taxonomy to identify common features in view of possible interpretation.

The actions enforced by ARPA-E in 2021 aroused our interest towards reconsidering with a new comprehensive approach to experimental data this longstanding issue that we have addressed almost at the same time with the paper published in Symmetry [23], already mentioned, where a theory based view point has been adopted with the discussion of experimental data limited to a small number of experiments.

The present paper, on the contrary, is not devoted to present a theoretical formalism, but to the systematic reconnaissance, deepening and classification of experimental evidences published by different authors. Most of the published papers adopt a fragmented and restricted approach in the sense not only that each one is concentrated on the single experimental layout utilized, but also that the results are very often presented as a confirmation of a theoretical interpretation suggested by that group of researchers specifically for their own results. Consequently, we considered useful, having put temporarily aside the theoretical assumptions adhering to which any single experiment was accomplished, to provide a synopsis in view of:

- commenting on the degree of fulfillment of high standards of design and implementation of the experiment, on the completeness of the details exposed and the extension of reproducibility achieved

- commenting on possible improvements of reproducibility

- identifying the possible presence of features common to several experiments that, beyond the details of each single experiment, may help in identifying some common action mechanisms to be considered shared among the different experimental approaches adopted.

Hence, the basic idea of the present paper, is first of all, the identification of a typology of cases each of them found representative of a variety of hundreds of experiments published in journals or in international congresses using the material environment where the experiment is conducted as a primary classification logic. The second step is the examination for each emblematic case, of a series of features as described in the following paragraph.

This work of collection, description, classification and comment of experiments already conducted may be a background providing support for the most valuable design (and possibly interpretation) of the next experiments promoted by the ARPA-E initiative mentioned above.

We are fully convinced that new experiments are necessary to exit from a sort of "suspension of judgment" on the issue of LENR; the outcome of the new experiments to be performed should be either to confirm, explain and possibly expand the indications acquired until now or to deny with convincing arguments (in primis through ad hoc experiments, in accordance with Popper's falsification principle) the credibility of the phenomena widely reported but until now controversial. To put the issue synthetically using old Greek words *aporìa* (contradiction) has generated *epoché* (suspension of judgment). The suspension should not last indefinitely (the question was raised more than 30 years ago with the claims by Fleishman and Ponds and it is still pending); *exegesis* (explanation and clarification) should be achieved promptly.

Moreover, we give indications in the direction of identifying the densification of energy (in all its different concrete realizations involving space and/or time dimensions) as a possible unifying concept applicable to all the emblematic cases considered.

In this sense, the conclusions reached in our previous article [23], already mentioned, where we indicated that the DST theory [22] allows to predict the outcomes resulting from a limited set of experiments, are extended to all the experiments examined in the present paper. This result can be considered a step forward overcoming a situation characterized by the circumstance that different theoretical models are in the available literature used for different experimental results.

We can therefore say, from the experimental results, that this theory provides a single and unambiguous interpretation for all anomalous nuclear physics results in all unconventional experiments. The unifying concept turns out to be the densification of energy in all its different concrete realizations. On the contrary, Authors of other experiments - and their related theoretical explanations - have always proposed phenomenological-level modeling hypotheses to interpret at least one experimental case without, however, being applicable to at least one other, different experimental case.

Our contribution aims to avoiding that the heritage associated with a wide engagement deployed for LENR during a long time may be disregarded or wasted both in terms of specific data gathered and in terms of methodologies acquired part of which might be utilized again with proper modifications.

## 2. METHODOLOGY ADOPTED TO ANALYZE EXPERIMENTAL EVIDENCES REPORTED IN LITERATURE

### 2.1 CHARACTERISTICS EXAMINED FOR THE CLASSIFICATION OF THE EXPERIMENTS

For the classification of the experiments the following three characteristics are examined:

- the type of observed experimental evidence which is considered "weird" (in the sense of being unexpected or even considered impossible according to generally accepted physical theories - anomalous being a synonym of "weird" in this respect) and the instrumentation adopted to this purpose
- the material matrix in which the evidence takes place
- the modalities through which the onset of the evidence is stimulated.

The evidence can be either microscopic (for instance detection of nuclei or particles not present before the onset of the phenomenon[3]) and/or macroscopic (such as either excess energy production in form of heat or pressure generated or visible modification in the appearance and structure of materials present in the experimental apparatus).

We call "experimental approach" the combination of item b. and item c. (for instance palladium electrodes in an electrolytic cell where voltage is used to stimulate the evidence expected).

The use of this classification is not only taxonomic. Hence suggestions can be obtained, in planning and performing further experimental activities, pertaining to the type of events to be searched and detected as well as to choices on the selection of instrumentation to be included in the experimental apparatus and even on the geometrical disposition of single detectors, since anisotropy characterizes the deployment of events resulting from this type of experiments. We underline that anisotropy of experimental outcomes can be the cause for lack of confirmation when a successful experiment is duplicated by a different group since minor differences in the layout of the experiment (for instance placement of detectors) my lead to miss the outcome looked for.

## 2.2 FIELD OF INVESTIGATION

The present synopsis is concentrated on experiments where transformation of matter is implied. Only "Nuclear non conventional phenomena" are considered while elsewhere (see for instance [23]) also non nuclear phenomena are addressed. Remaining in this specific field, we notice that the labelling "Low Energy Nuclear Reactions - LENR", is to some extent confusing since the expression "Nuclear reactions" is well established in physics to refer to the outcomes of nuclear forces as described by the Standard Model in the framework of which, the type of phenomena under investigation here are theroretically foreseen to be impossible. The expression "Nuclear non conventional phenomena" pregerred by us keeps the term nuclear since the modifications detected pertain to the nucleus, but drops the word "reaction" to be reserved to the consequences of the intervention of nuclear forces as described by the Standard Model. As a shortcut of "Nuclear non conventional phenomena" we recommend the use of "Nuclear metamorphosis" as well. It shoud be noticed that in physics the widest term ( i. e. the most generic) used in such situations is "interaction".

In the present work we do not consider the following two cases: catalytic nuclear fusion from muons[4] (mesic atom method) and ablation-catalyzed nuclear fusion generated by laser beams[5]. In fact, both cases do not encounter anomalies and are well encoded in the context of well-known applications of nonrelativistic quantum mechanics at the nuclear level (nuclear wave function superposition and consequent increase of tunnel probability).

The principal selection criteria have been semplicity of the layout and readability of the data describing both the experimental set up and the results substatiating the phenomenon under investigation. As already hinted, the main reasons for this choice have been on one side favouring duplication of experiments by other researchers, on the other side facilitating the identification of essential features common to the different experiments that could be useful to enlighten their nature.

## 2.2 TERMINOLOGY ADOPTED TO QUALIFY THE EVIDENCES REPORTED

As the title of this section underlines, the attention is focused on experimental evidences. By this term we mean what is found, highlighted, recorded and - that is decisive - measured in the experiment described in each paper; for instance, emission of particles quantitatively measured by means of adequate instrumentation properly operated and calibrated. The term evidence, to be interpreted as referring to a specific occasion, is not to be confused with the term phenomenon which refers to a systematic repetition of coherent evidences in given conditions.

---

[3] We use the term "evidence" to refer to a single outcome detected in a given type of experiment and the term "phenomenon" when there is a systematic repetition of coherent evidences in given conditions (see 2.3)

[4] A comprehensive and up-to-date overview of the open issues and prospects for further development of this approach can be found in the Imperial College doctoral dissertation (2018) [29] where the characteristics that a reactor based on this process should have are explored in depth. A subsequent paper outlines results and potential of an innovative method for muon production [30].

[5] See e.g the review papers [31], [32].

Since the different papers examined here are not based on a unique coherent terminology, it is considered useful to adopt the terminology described in the following table [33] to create a sort of hierarchy among situations with different level of reliability as a consequence of:

- sporadic or systematic occurrence

- degree of governability of the physical process characterizing the experiments

- multiplicity of research teams where experiments have been successful.

| TERM USED | | SITUATION OCCURRING AND INFORMATION ACQUIRED |
|---|---|---|
| Evidences | Sporadic evidence | Measurements indicate that some events take place (for instance the detecion of a particle) but the experimenter does not know when this may happen or not (i. e. control parameters are not identified clearly).<br>Even the *recipe* of what details may lead to the events looked for, is not clear. |
| | Reproducible evidence | The experimenter knows under what conditions the evidences appear.<br>A clear *protocol* indicates what are the controlling parameter and what value they should have for the evidences to appear (a correlation input output): an empirical model is attained. |
| Phenomenona | Occurrence of a phenomenon | A systematic repetition of coherent evidences in given conditions is obtained.<br>Phenomenological model are proposed to foresee the occurrence, for instance taking advantage of analogies with other phenomena. |
| | Interpretation of a phenomenon | Identification of well established physics-defined objects intervening and of their mutual interactions. |
| | Explanation of a phenomenon | Recurring to the asset of knowledge dealing with more general phenomena and laws, which can be used as a justification of the new phenomenon encountered. These knowledge can in some case be only hypothesized and expecting confirmation or refutation |

In synthesis, a set of experimental evidences must, as a prerequisite, be compliant with the generic concept of reproducibility[6] to be accepted as a proof of the existence of a physical phenomenon that is agreed to exist in nature. "Ideally, an experiment or analysis should be described in sufficient detail, so that other scientists with sufficient skills and means can follow the steps described in a published work and obtain the same results within the margins of experimental error" [33].

---

[6] There is not a widespread agreement on terminology used to describe different types of reproducibility; the following definitions might be useful.

*Repeatability* applies to a given context: the same results are obtained with stated precision in multiple trials, whenever the measurement is repeated, by the same team, using the same measurement system, under the same operating conditions, in the same location.

*Parametric repeatability* implies that the team has such a level of knowhow on managing the experiment that they can obtain coherent results, under different operating conditions on multiple trials.

*Replicability* corresponds to obtaining the same results by another team and elsewere, but under the same experimental conditions, rigidly following the original protocol.

*Reproducibility* indicates that other teams have fully understood the decisive features of the original test and that they can act on their own. This level of reproducibility by others acting autonomously is necessary to obtain consensus.

### *2.3 CHARACTERISTICS OF THE EVIDENCES RESULTING FROM THE EXPERIMENTS EXAMINED*

The experimental cases examined are, first of of all, classified according to the material  where the evidence takes place:

- solutions in gaseous or liquid environment

- metals of different composition and in different shape

- rocks and artificial chemical materials such as sequioxane.

### *2.3.1 Types of evidences encountered*

A distinction is introduced between:

- microscopic evidences consisting in: detection of neutrons or alpha particles not initially present; detection of nuclei not initially present; reduction of the quantity of an isotope initially present; reduction of radioactivity levels (gamma rays directly arising from a metamorphosis have not been detected, up to now.

- macroscpic evidences based on the appearance  of excess heat production and/or localized deformation of components of the experimental apparatus and/or modifications of radioactivity of a sample;

### *2.3.2 Identification of techniques used to stimulate the onset of the evidences*

The evidences appear only in the presence of actions that modify the conditions characterizing the environment. These actions can be classified as follows:

- compression via electricity

- compression via photons

- compression via ultrasounds with onset of cavitation

- compression via gas injection

- compression directly by shear (mechanical compression).

As anticipated in footnote 1 we use the term "compression" to indicate the operation activated by the experimenter; as such it is objective. We consider energy densification an inference of possible consequences of the operation on the status of the system.

## 3. COLLECTION OF THE INFORMATION PERTAINING TO THE EXPERIMENTS CONSIDERED

The information and the subsequent considerations formulated in this paper rely on a thorough examinations of the most significant information available in literature for each type of experiment that has provided evidences interpreted by the experimenters as a proof of the existence of phenomena in matter involving nucleus that can be called "weird" i.e. lacking until now widely accepted interpretation and explanation  (in several cases the use of the term phenomenon is questionable due to weakness in repeatability, see footnote 6 ).

### *3.1 SYNTHESIS DESCRIBING TYPES OF EXPERIMENTS AND EVIDENCES REPORTED*

In the following, an overall synthesis table is provided. For each experiment, the evidences detected are summarized in correspondence with the stimulation techniques used to generate such evidences. It should be noted that the meaning of the last column dealing with the modality leading to the energy densification will be explained later (see Sec. 3.3). In the first column of the tables reported below, the natural numbers, possibly followed by a letter, from 1. to 17, refer to the corresponding sheets in the Appendix.

*Part A. Solutions*

| Material where the evidence takes place | Compression via electricity | | | Compression via photons | | Compression directly by pressure | | | | Specific modality leading to the energy densification ramp |
| | Electric current | | Electric voltage | Photo stripped neutrons | Laser | | D gas insertion with pressure pulses | Shear | | |
| | Electric heating | In water solutions, D ions insertion in Pd | Pulsed discharges | Gamma irradiation of ErD$_3$ e TiD$_2$ | | Ultrasounds with cavitation | | No brittle fracture | With brittle fracture | |
| **Solutions** | | | | | | | | | | |
| Low pressure D2 gas with various types of cathode and Pd o W as anode **1.** | | | X-radiation having an energy nearly equal to the voltage applied to the discharge and energetic particle emission similar to deuterons having energy with peaks between 0.5 MeV and 3 MeV | | | | | | | Variations in space and volumes of the number of the force lines of the electric field (discharge is impulsive by definition) |
| Distilled water with Ti foil **2.a** | | | Glow discharge, new elements detected (B, Cu) and increment of isotope [48]Ti | | | | | | | Pulses of electric energy and associated concentration of electric charges |
| Distilled water with uranyl sulfate and Ti foil **2.b** | | | Distortion of the natural isotopic composition of uranium with consequent alteration of the secular equilibrium in uranium decay chain | | | | | | | Pulses of electric energy and associated concentration of electric charges |

| Distilled water with iron salts **3.** | | | | | | Detection of neutrons | | | | Ultrasonic cavitation |
| Distilled water with [228]Thorium **4.** | | | | | | Reduction of radioactivity levels and detection of new isotopes | | | | Ultrasonic cavitation |
| Nitric acid With [63]Ni **5.** | | | | | | Reduction of radioactivity levels and detection of new isotopes | | | | Ultrasonic cavitation |
| Deuterated acetone **6.** | | | | | | Detection of neutrons | | | | Ultrasonic cavitation without symmetrical spherical collapse |
| Palladium loaded with hydrogen in the presence of calcium carbonate **7.** | | Detection of neutrons | | | | | | | | Deformation of the electrodes that are the reactant |

**Part B. Metals Rocks and Artificial Compounds**

| Material where evidence takes place | Electric heating | In water solutions, D ions insertion in Pd | Pulsed discharges | Gamma irradiation of $ErD_3$ e $TiD_2$ | Laser | Ultrasounds with cavitation | D gas insertion with pressure pulses | No brittle fracture | With brittle fracture | Specific modality leading to the energy densification ramp |
|---|---|---|---|---|---|---|---|---|---|---|
| *Stimulation techniques* → | Compression via electricity | | | Compression via photons | | Compression via pressure | | Shear | | |
| | Electric current | | Electric voltage | Photo stripped neutrons | Laser | | D gas | Shear | | |
| **Metals** | | | | | | | | | | |
| Mercurio 8. | | | | | | Detection of new elements including rare earths | | | | Ultrasonic cavitation |
| Acciaio AISI 304 in aria 9.a | | | | | | Detection of neutrons | | | | Spherical symmetrical collapse of gas bubbles |
| Acciaio AISI 304 in aria 9.b | | | | | | Production of Cu with isotopic composition different from the natural one | | | | Spherical symmetrical collapse of gas bubbles |
| Acciaio AISI 304 in aria 10. | | | | | | | | Detection of alpha particles | | Time variation of pressure |
| Metallic powders in deuterium gas 11. | | | | | | | Production of excess heat | | | Compensation of pressure variation by pressure gauge inducing pressure shocks |
| Constantan in the presence of $H_2$ or $D_2$ 12. | Elements generation (C, O, Cl, Ca, and Zn) Excess heat production | | | | | | | | | Temperature increase |
| Palladium in the presence of $H_2$ or $D_2$ 13. | | | | | Elements generation | | | | | Temperature increase due to heating via He-Ne laser and excimer laser |
| Deuterated materials: $ErD_3$ and $TiD_2$ 14. | | | | Detection of photo-dissociation neutrons and claimed neutrons consistent with DT fusion producing $^3$He and a neutron | | | | | | Pulsed neutrons (generated by photo production) produce critical energy density throughout the volume and timing of the neutron bunches. |
| Mixture of Ni and $LiAlH_4$ 15. | Production of excess heat | | | | | | | | | Compensation, by current regulation, of temperature variation to obtain heat shocks |
| **Rocks** | | | | | | | | | | |
| Granites 16.a | | | | | | | | | Nuclear emissions: neutron detection | Fracture with cavitation through pressure shocks |
| Marbles 16.b | | | | | | | | No detection | | Fracture without cavitation (ineffective) |
| **Artificial compounds** | | | | | | | | | | |
| Silesquioxane 17. | | Excess heat (*) (*) Besides pulsed discharges, also photonic and ultrasound stimulations were used | | | | | | | | Pulsed time variation of the electric field. Ultrasonic cavitation |

### 3.2 DETAILED INFORMATION COLLECTED FOR EACH EXPERIMENT

In the Appendix, for each experiment examined, a descriptive identity card it is provided; this card is divided in two sections named "objective results" and "inferences" respectively.

### 3.2.1 "Objective" information for each experiment

The first set of information can be considered as "objective" which means that it was ascertained by the experimenters (it includes: what materials were present, what procedure were adopted and what was found, depending on what was looked for and on what instruments were used). Attention is focused on:

- Material where the evidence takes place
- Stimulation technique used to "trigger" the evidence
- Experimental evidences found
- Techniques employed to detect the evidences.

It should be stressed the issue of "what was looked for" and of "what instruments were used": for instance, in some cases, is presumable, in particular by comparison with the results of comparable experiments, that radiation could have been released but it was not detected because either it was not expected and consequently not looked for, or expected but not found because the necessary detection system was not installed.

### 3.2.2 Inferences for each experiment deriving from the interpretation of evidences

The second set of information, to be considered as inferences (in other words deductions, interpretations) formulated by us, includes:

- Estimated degree of description completeness and of reproducibility level
- Interaction environment
- Interaction agents
- Modality for energy densification
- Phenomenon type
- Microphysics Interpretation

The evaluation grid of repeatability of each experiment here introduced is not finalized to a judgment on the value of the paper, but only an indication of the viability of repetition with the available information. A four-value scale is used: Improvable, Sufficient, Good, Very good.

Interaction agent is the object, present in the apparatus, which, according to the interpretation adopted, participates to the onset of the events take place, eventually in synergy with secondary participants.
Interaction Environment indicates where the main Interaction agent is located and where potential secondary agents come from.

Experiments providing microscopic evidences of *Nuclear Metamorphosis* (concerning element(s) production or variation from one starting element to another or isotopic variations that occur in an appropriate and appropriately stressed environment without the use of radiation)[7] are categorized as follows:

---

[7] As already mentioned "nuclear" is used here in the sense that a nucleus is transformed, but the adjective indicates "where the transformations occur" not the intervention of nuclear forces as commonly defined

(*Reduction of radioactivity*)[8]
From a substance containing at least one radioactive element or one radionuclide stable elements are produced as a result of transformation of the starting radionuclides and reduction of the activity level of the starting substance from that before the treatment. Possible application area: Removal of radioactivity from nuclear waste.

*Isotopic change*
From a multi-isotopic starting element, stable isotopes are produced with alteration of the natural isotopic abundances. Possible application area: Removal of radioactivity from nuclear waste as demonstrated in experiments conducted with Thorium and Nickel (see. in the Appendix experiment type 4 and 5 respectively).

*Production of elements*.
From a stable starting element, stable elements previously not present in the sample (normally lighter than the starting one) are produced
Possible application area: Production of valuable elements, for example rare earths or helium.

*Nuclear emissions* [9]
Occurrence of neutrons or α particles that are NOT α radiation from nuclear decay.
Possible application area: Neutron or α particle source productions; power generation.

This classification aims also at justifying the use of the term phenomenon (see 2.2) since a certain type of evidences are repeated in different conditions.

Microscopic interpretation is a tentative description of the phenomenon in terms of interactions among well-established physical objects, such as particles and nuclei.

In the analysis of the literature, additional information, besides what is listed above, was acquired, when available, dealing with:
-   experimental conduct ("how it was proceeded" also in view of reproducibility considering the parameters influencing the outcome
-   the possibility of adjusting and controlling the process)
-   additional outcomes during the experiments.

This information contributes, as source of clues, to substantiate the inferences deduced. Among the additional outcomes one should mention:
-   Production of air bubbles
-   Production of debris
-   Deformation of electrodes
-   Light emission
-   Presence of hysteresis phenomena
-   Pulse nature of the phenomenon
-   Occurrence of micro explosions
-   Effects of positioning of detectors
-   Effects of geometry of the stimulator device (electrode or sonotrode) or of its surface treatment.

By inter-comparing the occurring of these clues, suggestions can also be derived about additional instrumentation to be used to detect other evidences in further, hopefully conclusive experiments.

---

[8] In a previous work (Refs. 4.1, 4.2 cited in the card dealing with the experiments of type 4 in Appendix) this phenomenon of reducing the radioactivity level of the sample subjected to sonication has been called "neutralization".

[9] The distinction between Nuclear Emission and Nuclear Metamorphosis is valid if one remains at the level of found evidence. Specifically: neutron and/or α emission is to be assumed to involve that some nucleus has been modified in that it "lost" the emitted neutron or particle, unless one imagines no conservation of the total number of hadrons. Emission involves nuclear metamorphosis while metamorphosis does not necessarily involve emission.

Even though, in many cases, the additional data mentioned above have been valuable to provide clues to formulate the inferences, they were not reported in the data sheets for lack of space. Interested readers may have access to these data looking at the References mentioned for each experiment.

We underline that, for each experiment type, three different levels of details are made available, corresponding respectively to:
- the synthesis presented in section 3.1
- the dedicated data sheet reported in Appendix
- the bibliographic references to the original papers describing each experiment mentioned in each data sheet.

## 3.3 AN INTERPRETATION OF THE ENERGY DENSIFICATION MECHANISMS

### 3.3.1 The process of energy densification

We call energy densification the process of reaching energy densities that are critical - in space and time - for the occurrence of the phenomenon[10], which may have entirely new and peculiar characteristics, even to the point of apparently contradicting well-known and acquired laws, such as, for example, the conservation of total energy and the second principle of thermodynamics. But such violations are only apparent and are resolved in the context of DST theory [23] [22], which treats spacetime as an energy-dependent elastic medium and contextually energy reservoir, without invoking any concept involving the vacuum state.

The existence of energy thresholds is found to separate, for each interaction, the flat metric part from the deformed metric part. The phenomena occurring in the deformed part of the interaction metric are governed by the energy density in the spacetime (volume [34] and time interval [35] and Chap. 16, par. 16.3.1 p. 242 – 245, par. 16.3.5  p. 249 – 251 of [22]).

From the mathematical standpoint, the energy E has to be considered as a dynamic variable, because it specifies the dynamic behavior of the process under consideration, thus providing, through the metric coefficients, a dynamic map in the energy range of interest of the interaction ruling a given process, with regards to the energy density in the space volume [34]. At the experimental and practical level, the outcome of energy densification results in reaching the values of critical pressure, i.e., energy per unit volume, and critical power, i.e., energy in the interval of time defined as unit, at which the phenomenon is triggered and produces measurable effects. With respect to the energy density in the time variable, the power density for piezonuclear reactions was also estimated, as an order of magnitude, starting from the DST theory [35].

### 3.3.2 Modes to obtain stimulation in different configurations

In particular, in the case of a liquid medium, densification occurs by employing ultrasound to induce critical conditions in phase space with a mechanism that can be traced back to cavitation,[11] which gives conditions for energy thickening. In metals, it can be imagined that the confinement function can be fulfilled by Ridolfi cavities and the ultrasonic wave can be generated by fracture propagation[12].  In rocks, it can be imagined that Ridolfi cavities would intervene, but only in the presence of brittle fracture and subsequent pressure wave. More on this subject is set forth in reference [21] chapter 16 paragraph 16.3 pages 242-251. There it is reported that the orders of magnitude of cell sizes[13] are those in the range of (4 to 8) microns for space and microsecond for time respectively.

---

[10] The label "quadridimensional densification" is suggested to express that both energy density in the space volume and energy density in the time interval must be considered.

[11] Cavitation arises when a plane wave of pressure of given wavelength impinges on a gaseous bubble internal to matter if the diameter of the bubble is much less than the wavelength: by Pascal's principle it produces a spherical-symmetric collapse in that bubble.

[12] It is well known that there are very high stress and therefore energy concentrations at the tip of a propagating crack

[13] We mean micro cavities, with diameters between 1 and 10 microns, in the material rather than cells in the sense of the Liouville space, nor commonly known voids in metallurgy, e.g., the well-known Kirkendall voids (40-60 microns).

In the case of a gas pressure on metal powders, it is the gas pressure that acts directly on the matter producing the acceleration in phase space that leads to the critical cell and thus to the conditions of nuclear metamorphosis of matter.

In the case of a massive material subjected to hydrogen loading, mechanical stresses arise. They can change the microstructure of the material until local microfractures develop. The latter can occasionally give rise to an accelerated increase of the pressure, thus leading to the microcavity reaching critical conditions, and hence to the nuclear metamorphosis occurring into corresponding the phase space cell. Since the development of such local fractures is occasional and cannot be linked in a predictable way with hydrogen loading, the critical conditions in phase space are difficult to reproduce.

For a solid material directly subjected to mechanical stress, up to a critical load, its overcoming by fracture is produced. In the event that the fracture is "brittle" there is an acceleration leading to nuclear metamorphosis into the phase space inside the critical cell. Should the fracture be "ductile", the acceleration is insufficient and never reaches the critical conditions in the cell, needed to nuclear metamorphosis.

To the purpose of obtaining an energy densification method useful to generate phenomena of nuclear metamorphosis in condensed matter, in particular in solid matter, it's possible to consider also the class of processes consisting in "loading" light elements into the bulk of a solid matrix also of crystalline type; this loading process has been widely used in experiments conducted in the framework of experimental activities dealing with the so called "cold fusion" for instance using hydrogen or deuterium as gases to be introduced in a metal bulk material. Nevertheless, such a process is highly unpredictable with respect to the aim of reaching the critical energy density necessary to obtain nuclear metamorphosis. Even if the deformation of the material undergoing loading can supply clues on the energy densification process, any single loading operation has its own dynamic and is characterized by its own history that do not necessarily ensure a highly reliable reproducibility when targeting the onset of nuclear metamorphosis; this is shown by the discording results to be registered in experimental layouts that appear to be overlapping.

## 4. COMMENTS AND CONTRIBUTIONS FOR FURTHER STEPS TOWARDS SCENARIO CLARIFICATION

### 4.1 ISSUES AND TOPICS SHOWING WEAKNESSES IN THE EXECUTION OF RESEARCH PROGRAMS
### Difficulties in replicating in other laboratories has been the Achilles' heel of this research area

As already noticed, a problem displayed by the existing literature on this topic, is that the experimenter himself sometimes fails to reproduce what he has achieved. Moreover, when the experiment is repeated elsewhere there are difficulties in obtaining results coherent with the outcomes of the original experiment (see section 2.2 above). This can be due to the adoption of experimental procedures which are incorrect with respect to those adopted in the experiment to be repeated.

An important aspect, to be kept in mind, is that in the designing and performing the considered experiments, there has been a remarkable lack of modeling support, from which to draw guidance on what to look for. Additional problems can arise, as a consequence of the intrinsic characteristics of the phenomena considered here, such as anisotropy, asymmetry, asynchrony, inhomogeneity of emission that lead to specific difficulties in defining the experimental layout that are not universally perceived by experimenters.

Another sensitive issue is related to the delicacy of instrumentation with respect to false signal geometries and background intensities. The decisive impact of the interaction between theoretical models and experiments, within this context, shows the existence of two opposing errors: total lack of a theoretical model to drive the design of the experiment; inability to capture unexpected evidence.

Some role is also played by the equivocal and unsupported terminology, such as, for example piezonuclear metamorphosis, LENR, cold fusion, a non-unified terminology to denote the same set of experiments, with different, sometimes provisional names. There has been historically, in the development of the field, closure and opposition between research groups, as well as interference between sharing scientific heritage and protecting intellectual heritage [2], [4], [7], [8].

## 4.2 SOME FEATURES SIMULTANEOUSLY PRESENT IN ALL THE EXPERIMENTS

### 4.2.1 A possible reversal of the most common story telling of LENR

The most common reading of a LENR mechanism is that light nuclei properly stimulated and in the presence of a properly chosen piece of matter (most frequently a metal properly treated) are allowed to react with each other in contradiction with consolidated physics laws. Looking at the following table (which is simply a collection of data reported in the second section of the sheets describing the experiments considered) an alternative exposition appears to be consistent with experimental results: under proper conditions nuclei of elements having a high value of binding energy per nucleon when are properly stimulated by nuclei of light elements undergo phenomena of nuclear metamorphosis. Among consequences of such a formulation one might mention possible hints on the most promising choice of materials to be utilized for additional experiments. In the following table we mention only experiments where new isotopes of already present nuclei, or new nuclei altogether are produced.

| Experiment type | Interaction Environment | Interaction agent | Products, as reported by Experimenters |
|---|---|---|---|
| 1 | Deuterium gas and or deuterium in anode metal lattice | Deuterium and /or atoms in metals | Several isotopes not present before |
| 2 a | Distilled water | Elements present in Ti foils | Several isotopes not present before |
| 2 b | Titanium in water containing uranium salts | Elements present in Ti foils and uranium | Several isotopes not present before |
| 3 | Water | $^{56}$Fe | Several isotopes not present before |
| 4 | Water | $^{228}$Th | Several isotopes not present before |
| 5 | Nitric acid solution | $^{63}$ Ni nitrate | Several isotopes not present before |
| 6 | Deuterated acetone | Deuterated acetone | Several isotopes not present before |
| 7 | Palladium electrodes in water solution of electrolytic salts | Elements present in Pd metal | Several isotopes not present before |
| 8 | Mercury | Mercury | Several isotopes not present before |
| 9 a | Air present in the internal microcavities of AISI 304 austenitic steel | Elements present in AISI 304 austenitic steel | Several isotopes not present before |
| 9 b | Air present in the internal microcavities of the material | Elements present in AISI 304 austenitic steel | Copper with isotopic composition different from that of natural copper |
| 10 | AISI 304 austenitic steel | Elements present in AISI 304 austenitic steel | Several isotopes not present before |
| 11 | Deuterium gas | Metal powders | Several isotopes not present before |
| 12 | Constantan lattice deformed, at the surface of the wire, by the presence of deuterium or hydrogen | All elements present | Several isotopes not present before |
| 13 | Pd lattice deformed by the presence of deuterium or hydrogen | All metals presents | Several nuclides not present before |
| 14 | Deuterated Erbium and deuterated Ti | Deuterium and possibly erbium and titanium as well | Several nuclides not present before |
| 15 | Hydrogen atmosphere | Nickel hydride | Several nuclides not present before |
| 16 a | Air but not free air; metamorphosis takes place in the air enclosed in cavities contained in granites | Granites with iron in air | Several nuclides not present before |

***4.2.2 A hypothesis on the mechanism of action applicable across all configurations: densification***

As seen in section 3.3., in order to enunciate a unified interpretation of the above findings, it has been useful to resort to language analogous to that employed in the phase space formalism. Indeed, as we have seen, we defined "cells" as appropriate portions of space identified by defined intervals of the spatial coordinates and the time coordinate, and introduced the concept of energy density, as the amount of energy present per unit volume of the (just defined) cell. The external action consists of interventions to increase this density in the cells, until a critical threshold is reached at which nuclear metamorphosis phenomena with nucleosynthesis and nucleolysis allowed by generalized Lorentz invariance are triggered between nuclei. Since densification occurs in a critical volume or critical time interval or both, by analogy with phase space, a cell that has reached critical conditions is called a "critical cell." It can be viewed as an energy confinement space.

In the presence of pressure wave-induced cavitation, one can imagine that bubbles serve the function of confinement space. The unifying element of the different experiments considered in the reconnaissance can be defined thus with regard to the intervention leading to the phenomenon under study: A steep enough ramp of thickening (densification) of the energy in the time interval and in the volume of active space occurs, until a sufficiently high value of energy concentration is reached (like a threshold to be exceeded).

Finally, we underline a classification of the experiments examined, assuming as a criterion the way adopted in attaining the energy densification.

| Experiment identification | Way adopted in attaining the energy densification (densification type) |
|---|---|
| 1 | A. Variation in volume and time of the number of the force line of the electric field |
| 2.a, 2.b, 3, 4, 5, 6, 8, 9.a, 9.b, 17 | B. Pulses of electric energy and concentration of the electric charges |
| 7, 10, 11, 16.a, 16.b, 17 | C. Pressure variations |
| 12, 13, 15 | D. Temperature variations |
| 14 | E. Action by pulsed neutrons |

As a complementary information we remind that a thorough treatment of densification thresholds from a theoretical view point can be found in Chap. 16, par. 16.3.1 p. 242 – 245, par. 16.3.5 p. 249 – 251 of Reference [22].

## 5. CONCLUDING CONSIDERATIONS AND SUGGESTIONS

We have conducted a comprehensive selection, classification and comparative analysis of what we consider the most significant experiments dealing with LENR reported in literature and have derived a broad reconnaissance of the experimental results aiming at the construction of a systematic phenomenology. The unifying concept has turned out to be the densification of energy in all its different concrete realizations. We have observed that the experimental outcomes are in accordance with theoretical predictions of a single theory for all experiments, the DST theory which was the object of a recent paper published by us on Symmetry [23]. To be noticed that, on the contrary, authors of other experiments - and their related theoretical explanations - have always proposed phenomenological-level modeling hypotheses to interpret one experimental configuration case or, in some instances, only several experimental configurations.

Even some of the reviews of experiments and their corresponding evidences are marred by being biased and too much aimed at proving a prejudicial thesis, with the characteristics of being an overly specific phenomenological level modeling. This interpretive landscape highlights a proliferation of uncoordinated experimental attempts lacking reasonable mutual integration at the level of results. In addition, many experimental attempts used multiple tools to stress the matter in order to stimulate the occurrence of events; such a situation jeopardized a clear, hierarchical separation of the effects of each stress.

In essence, little use was made, in most of the literature so far, of the ability to link cause to effect. Thus, the first of our purposes here has been that of providing an opportunity for the international scientific community to move from the level of individual evidence to that of identifying elements for a properly articulated and hopefully shared view of the phenomena considered here. A transition crossed in similar circumstances in the history of physics for example at the discovery of the neutron and the discovery of nuclear fission.

The second purpose is to promote a research program that brings clarity with an overall assumption of responsibility by the scientific community, aimed at working together for a systematization of the various theoretical contributions made by different research groups. In particular, it is noted that the inference of the existence of a mechanism common to all experiments consisting of an energy densification ramp is coherent with the theory of deformed space time leading to the identification of energy thresholds, which are used to calculate the critical densities triggering the phenomena.

From this viewpoint, it is useful to distinguish the two concepts of the critical density of energy and the energy densification. The latter is the practical and technical way to obtain the critical density of energy in each experimental setup, with respect to the phenomena to be investigated. For this reason, we refer to energy densification in each identity card given in Appendix, concerning the different experiments we examined and listed.

It is not coherent with a systematic investigation of physical phenomena that, after thirty years, questions are still pending on the existence, significance and impact of LENR to which the most acknowledged physical theories are not capable of giving an answer. Even a definitive demonstration that all these experiments have decisive faults would be preferable than leaving the issue unaccounted for. Welcoming the decision of ARPA-E to launch an initiative to investigate in a coordinated manner with specific tasks asigned the issue of LENR, we consider that the organized review of previous experiments on this subject and the other considerations presented in the present paper can give a useful background helping design, execution and interpretation of the next experiments promoted by the ARPA-E initiative mentioned above.

## *ACKNOWLEDGMENTS*

The authors gratefully tank Maurizio Maggiore, Policy Officer at Directorate-General for Research and Innovation of the European Commission, for useful discussions.

*APPENDIX: IDENTITY CARD FOR EACH EXPERIMENT*

The identity cards of the experiments are summarized in the sheets named "experiment type" from 1. to 17.; sometimes, when needed, there are subtypes denoted by letters, e.g. 1.a. The acronyms of the detectors and the techniques of analysis related to the experiments reported in the sheets are given hereafter.

### Detectors of neutrons

Thermodynamical detector BTI (Bubble Technology Industry) using hydrocarbon , name: Defender, DefenderXL

Electronic detector using BF3 (boron trifluoride)

Electronic detector using He3 (Helium-3)

Photographic imaging by PADC  (polyallyl diglycol carbonate) type CR39- Boric Acid

### Detectors of charged particles:

Photographic imaging by PADC  (Polyallyl Diglycol Carbonate) type CR39

Electronic detector using ZnS (Ag) (Zinc Sulphide with Silver doping)

Silicon barrier detector SBD

### Analysis techniques:

HF Detector (Gas Analyzer of hydrogen fluoride) , name : GASTiger2000 HF 3

X Ray Spectroscopy by NaI (sodium iodine)

XRF (X Ray Fluorescence)

ICP-OES (Induced Coupled Plasma - Optical Emission Spectroscopy)

ICP-MS (Induced Coupled Plasma - Mass Spectrometry)

SEM (Scanning Electron Microscope)

ESEM-EDS (Enviromental Scanning Electron Microscope - Energy Dispersion Spectroscopy)

BSE imaging (Back Scattered Electrons)

INAA (Instrumental Neutron Activation Analysis) using gamma ray spectroscopy

***EXPERIMENT TYPE 1.***

***Objective results***

| | |
|---|---|
| MATERIAL | Low pressure $D_2$ gas with various types of cathode and Pd or W as anode |
| STIMULATION TECHNIQUE | Electric discharge in gaseous or vapor atmosphere |
| EXPERIMENTAL EVIDENCES | X-radiation having an energy nearly equal to the voltage applied to the discharge and energetic particle emission similar to deuterons having energy with peaks between 0.5 and 3 MeV |
| TECHNIQUES TO DETECT THE EVIDENCES | Silicon barrier detector (SBD) and Geiger Muller counter |

***Inferences***

| | |
|---|---|
| ESTIMATED DEGREE OF DESCRIPTION COMPLETENESS & OF REPRODUCIBILITY LEVEL | Reported as very good and with 100 % reproducibility, but documentation is undisclosed at present |
| INTERACTION ENVIRONMENT | Deuterium gas and or deuterium in anode metal lattice |
| INTERACTION AGENT | Deuterium and /or atoms in metals |
| MODALITY FOR ENERGY DENSIFICATION | Variations in space and volumes of the number of the force lines of the electric field (discharge is impulsive by definition) |
| PHENOMENON TYPE | Emission of energetic particles |
| MICROPHYSICS INTERPRETATION | In the presence of nuclei of hydrogen $^{1}$H, deuterium $^{2}$D, and of various metal in the electrodes under proper stimulation, several isotopes before not present are produced in coherence with the Baryons Conservation law pertaining to nuclear reactions which is considered to be valid also in this case. |

***EXPERIMENT TYPE 2. a***

***Objective results***

| | |
|---|---|
| MATERIAL | Titanium in distilled water |
| STIMULATION TECHNIQUE | Impulsive electric discharge and fast rupture of one Ti electrode shaped as a foil |
| EXPERIMENTAL EVIDENCES | Glow discharge, new elements detected (B, Cu) and increment of isotope Ti 48 |
| TECHNIQUES TO DETECT THE EVIDENCES | OES, XRF, ICP-MS |

***Inferences***

| | |
|---|---|
| ESTIMATED DEGREE OF DESCRIPTION COMPLETENESS & OF REPRODUCIBILITY LEVEL | Good |
| INTERACTION ENVIRONMENT | Distilled water |
| INTERACTION AGENT | Elements present in Ti foils |
| MODALITY FOR ENERGY DENSIFICATION | Pulses of electric energy and associated concentration of electric charges |
| PHENOMENON TYPE | Nuclear Metamorphosis: Production of elements and nuclides. |
| MICROPHYSICS INTERPRETATION | In the presence of nuclei present in Ti foils, under proper stimulation, several isotopes before not present are produced together with nuclear particles in absence of gamma emissions and in coherence with the Baryon Number Conservation law pertaining to nuclear reactions (which is considered to be valid also in this case). |

***EXPERIMENT TYPE 2. b***

***Objective results***

| | |
|---|---|
| MATERIAL | Solution of uranyl sulfate in distilled water |
| STIMULATION TECHNIQUE | Impulsive electric discharge and fast rupture of one Ti electrode shaped as a foil |
| EXPERIMENTAL EVIDENCES | Distortion of the natural isotopic composition of uranium with consequent alteration of the secular equilibrium in uranium decay chain. |
| TECHNIQUES TO DETECT THE EVIDENCES | α, β, γ-spectrometry and mass-spectrometry |

***References***

2.2 *Study of the Electric Explosion of Titanium Foils in Uranium Salts,* Leonid I. Urutskoev, Dmitry V. Filippov, *J.* Mod. Phys., 2010, 1, 226-235

***Inferences***

| | |
|---|---|
| ESTIMATED DEGREE OF DESCRIPTION COMPLETENESS & OF REPRODUCIBILITY LEVEL | Good |
| INTERACTION ENVIRONMENT | Titanium in water containing uranium salts |
| INTERACTION AGENT | Elements present in Ti foils and uranium |
| MODALITY FOR ENERGY DENSIFICATION | Pulses of electric energy and associated concentration of electric charges |
| PHENOMENON TYPE | Nuclear Metamorphosis: Production |
| MICROPHYSICS INTERPRETATION | In the presence of nuclei present in Ti foils and of uranium, under proper stimulation, several isotopes before not present are produced together with nuclear particles in absence of gamma emissions and in coherence with the Baryon Number Conservation law pertaining to nuclear reactions (which is considered to be valid also in this case). |

**EXPERIMENT TYPE 3.**

**Objective results**

| | |
|---|---|
| MATERIAL | Water with iron |
| STIMULATION TECHNIQUE | Ultrasound 20 Khz |
| EXPERIMENTAL EVIDENCES | Emission of neutrons |
| TECHNIQUES TO DETECT THE EVIDENCES | Thermodynamic BTI Defender, DefenderXL ; Electronic BF3 |
| | Photographic PADC CR39-Boric Acid |

**Inferences**

| | |
|---|---|
| ESTIMATED DEGREE OF DESCRIPTION COMPLETENESS & OF REPRODUCIBILITY LEVEL | Good |
| INTERACTION ENVIRONMENT | Water |
| INTERACTION AGENT | $^{56}$Fe |
| MODALITY FOR ENERGY DENSIFICATION | Ultrasonic cavitation |
| PHENOMENON TYPE | Nuclear Emissions |
| MICROPHYSICS INTERPRETATION | In the presence of nuclei of iron under proper stimulation, several isotopes before not present are produced together with nuclear particles in absence of gamma emissions and in coherence with the Baryons Conservation law pertaining to nuclear reactions (which is considered to be valid also in this case). |

$^{56}$Fe 91,7 % $^{54}$Fe 5,82 %, $^{57}$Fe 2,19 %, $^{58}$Fe 0,28%.

***EXPERIMENT TYPE 4.***

***Objective results***

| | |
|---|---|
| MATERIAL | Water with $^{228}$Th |
| STIMULATION TECHNIQUE | Ultrasounds 20 Khz |
| EXPERIMENTAL EVIDENCES | Reduction in radioactivity of the original sample. Reduction in the presence of original radionuclides |
| TECHNIQUES TO DETECT THE EVIDENCES | ICP-MS, Photographic PADC CR39 |

***Inferences***

| | |
|---|---|
| ESTIMATED DEGREE OF DESCRIPTION COMPLETENESS & OF REPRODUCIBILITY LEVEL | Sufficient |
| INTERACTION ENVIRONMENT | Water |
| INTERACTION AGENT | $^{228}$Th |
| MODALITY FOR ENERGY DENSIFICATION | Ultrasonic cavitation |
| PHENOMENON TYPE | Nuclear metamorphosis: neutralization of radioactivity |
| MICROPHYSICS INTERPRETATION | In the presence of nuclei of $^{228}$Th under proper stimulation, several isotopes before not present are produced together with nuclear particles in absence of gamma emissions and in coherence with the Baryon Number Conservation law pertaining to nuclear reactions (which is considered to be valid also in this case). This is an application of nuclear metamorphosis with transformation of radioactive substances, an application that has been called neutralisation of radioactivity. |

**EXPERIMENT TYPE 5.**

**Objective results**

| | |
|---|---|
| MATERIAL | Solution of nitric acid with nitrate of $^{63}$Ni |
| STIMULATION TECHNIQUE | Ultrasounds 35 Khz |
| EXPERIMENTAL EVIDENCES . | Reduction in radioactivity of the original sample. Detection of new isotopes not present in the original sample: First group: $^{60}$Ni, $^{59}$Co, $^{11}$Be, $^{9}$Be, $^{7}$Li . Second group: $^{23}$Na, $^{39}$K, $^{44}$Ca, $^{51}$V, $^{69}$Ga, $^{75}$As, $^{77}$Se, $^{85}$Rb, $^{88}$Sr, $^{95}$Mo, $^{107}$Ag, $^{111}$Cd, $^{115}$In, $^{118}$Sn, 1$^{121}$Sb, $^{133}$Cs, $^{137}$Ba, $^{139}$La, $^{140}$Ce, $^{205}$Tl, $^{208}$Pb, $^{209}$Pb, $^{238}$U . |

**Inferences**

| | |
|---|---|
| ESTIMATED DEGREE OF DESCRIPTION COMPLETENESS & OF REPRODUCIBILITY LEVEL | Good |
| INTERACTION ENVIRONMENT | Nitric acid solution |
| INTERACTION AGENT | $^{63}$Ni nitrate |
| MODALITY FOR ENERGY DENSIFICATION | Ultrasonic cavitation |
| PHENOMENON TYPE | Nuclear metamorphosis: neutralization of radioactivity |
| MICROPHYSICS INTERPRETATION | In the presence of nuclei of $^{63}$Ni under proper stimulation, several isotopes before not present are produced together with nuclear particles in absence of gamma emissions and in coherence with the Baryon Number Conservation law pertaining to nuclear reactions (which is considered to be valid also in this case). The nuclides of the first group derive from the metamorphosis of $^{63}$Ni; the nuclides of the second group derive from the metamorphosis of other elements that are present. |

*EXPERIMENT TYPE 6.*

*Objective results*

| MATERIAL | Deuterated acetone |
| --- | --- |
| STIMULATION TECHNIQUE | Variable-frequency ultrasound |
| EXPERIMENTAL EVIDENCES | Neutron Emission |
| TECHNIQUES TO DETECT THE EVIDENCES | Electronic $^3$He |

*Inferences*

| ESTIMATED DEGREE OF DESCRIPTION COMPLETENESS & OF REPRODUCIBILITY LEVEL | Improvable |
| --- | --- |
| INTERACTION ENVIRONMENT | Deuterated acetone |
| INTERACTION AGENT REAGENT | Deuterated acetone |
| MODALITY FOR ENERGY DENSIFICATION | Ultrasonic cavitation without symmetrical spherical collapse |
| PHENOMENON TYPE | Nuclear emissions |
| MICROPHYSICS INTERPRETATION | In the presence of nuclei of Deuterium under proper stimulation several isotopes before not present are produced together with nuclear particles in absence of gamma emissions and in coherence with the Baryon Number Conservation law pertaining to nuclear reactions (which is considered to be valid also in this case). |

***EXPERIMENT TYPE 7.***

***Objective results***

| | |
|---|---|
| MATERIAL | Palladium electrodes charged with hydrogen or deuterium and electrolyte salts (calcium carbonate) |
| STIMULATION TECHNIQUE | Electrolytic current |
| EXPERIMENTAL EVIDENCES | Neutron emission |
| TECHNIQUES TO DETECT THE EVIDENCES | Photographic PADC CR39 |

***References***

7.1 *Comparison of Pd/D co-deposition and DT neutron generated triple track observed in CR-39 detectors,* P.A. Mosier-Boss et al. European Physical Journal  of Applied Physics  51,2, 20901-20911 (2010)

7.2  *Condensed Matter Nuclear Science Using Pd/D Co-Deposition* P.A. Mosier-Boss L. Forsley Research Gate 2015 https://www.researchgate.net/publication/283569283_Condensed_Matter_Nuclear_Science_Using_PdD_Co-Deposition

***Inferences***

| | |
|---|---|
| ESTIMATED DEGREE OF DESCRIPTION COMPLETENESS & OF REPRODUCIBILITY LEVEL | Sufficient |
| INTERACTION ENVIRONMENT | Palladium electrodes in water solution of electrolytic salts |
| INTERACTION AGENT | Elements present in Palladium metal |
| MODALITY FOR ENERGY DENSIFICATION | Deformation of the electrodes that are the reactant |
| PHENOMENON TYPE | Nuclear emissions |
| MICROPHYSICS INTERPRETATION | In the presence of hydrogen in the palladium bulk under proper stimulation, several isotopes before not present are produced together with nuclear particles in absence of gamma  emissions  and in coherence  with  the  Baryon Number Conservation  law pertaining to nuclear reactions (which is considered to be valid also in this case). |

*EXPERIMENT TYPE 8.*

*Objective results*

| MATERIAL | Liquid mercury |
| --- | --- |
| STIMULATION TECHNIQUE | 20 Khz and 35 Khz ultrasounds |
| EXPERIMENTAL EVIDENCES | Presence of elements absent before the stimulus, among which some rare earths are revealed: $^{89}$Y, $^{138}$Ce, $^{151}$Eu, $^{152}$Gd,$^{158}$Gd, $^{174}$Yb, $^{176}$Lu |
| TECHNIQUES TO DETECT THE EVIDENCES | ICP-OES, ICP-MS, SEM, ESEM-EDS, XRF, INAA |

*Inferences*

| | |
| --- | --- |
| ESTIMATED DEGREE OF DESCRIPTION COMPLETENESS & OF REPRODUCIBILITY LEVEL | Very good |
| INTERACTION ENVIRONMENT | Mercury |
| INTERACTION AGENT | Mercury |
| MODALITY FOR ENERGY DENSIFICATION | Ultrasonic cavitation |
| PHENOMENON TYPE | Nuclear Metamorphosis: Production of elements and nuclides. |
| MICROPHYSICS INTERPRETATION | In the presence of nuclei of mercury under proper stimulation, several isotopes not present before are produced together with nuclear particles in absence of gamma emissions and in  coherence  with  the  Baryon Number Conservation law pertaining to nuclear reactions (which is considered to be valid also in this case). |

**EXPERIMENT TYPE 9.a (investigation focused on the environment surrounding the sample)**

**Objective results**

| | |
|---|---|
| MATERIAL | AISI 304 austenitic steel in air |
| STIMULATION TECHNIQUE | 20 kHz ultrasounds |
| EXPERIMENTAL EVIDENCES | Detection of neutrons |
| TECHNIQUES TO DETECT THE EVIDENCES | Electronic $^3$He; Photographic PADC CR39-Boric acid |

**References**

9.1 *Piezonuclear neutrons from iron*, F. Cardone, R. Mignani, M. Monti, A. Petrucci, V. Sala, Modern Physics Letters A, Vol. 27, 18, 1250102, 2012

9.2 *Violation of local Lorentz invariance for Deformed Space-Time neutron emissions* F. Cardone, G. Cherubini, M. Lammardo, R. Mignani The European Physical Journal-Plus 130, 35, 2015

9.3 *Energy spectra and fluence of the neutrons produced in Deformed Space-Time conditions* F. Cardone, A. Rosada, Modern Physics Letters B 30, 28, 16503461-7, 2016

9.4 *Deformed Space-Time neutrons: spectra and detection,* F. Cardone, G. Cherubini, A. Rosada, Journal of Advanced Physics 7, 1, 81-87, 2018

**Inferences**

| | |
|---|---|
| ESTIMATED DEGREE OF DESCRIPTION COMPLETENESS & OF REPRODUCIBILITY LEVEL | Good |
| INTERACTION ENVIRONMENT | Air present in the internal microcavities of AISI 304 austenitic steel |
| INTERACTION AGENT | Elements present in AISI 304 austenitic steel |
| MODALITY FOR ENERGY DENSIFICATION | Ultrasonic cavitation |
| PHENOMENON TYPE | Nuclear emissions |
| MICROPHYSICS INTERPRETATION | In the presence of nuclei of elements present in AISI 304 austenitic steel under proper stimulation, several isotopes not present before are produced together with nuclear particles, in absence of gamma emissions and in coherence with the Baryon Number Conservation law pertaining to nuclear reactions (which is considered to be valid also in this case). |

**EXPERIMENT TYPE 9.b (investigation focused on the sample)**

**Objective results**

| | |
|---|---|
| MATERIAL | AISI 304 austenitic steel in air |
| STIMULATION TECHNIQUE | 20 kHz ultrasounds |
| EXPERIMENTAL EVIDENCES | Production of copper having isotopic composition different from that of natural copper |
| TECHNIQUES TO DETECT THE EVIDENCES | ESEM-EDS, BSE-imaging, INAA |

**Inferences**

| | |
|---|---|
| ESTIMATED DEGREE OF DESCRIPTION COMPLETENESS & OF REPRODUCIBILITY LEVEL | Good |
| INTERACTION ENVIRONMENT | Air present in the internal microcavities of the material |
| INTERACTION AGENT | Elements present in AISI 304 austenitic steel |
| MODALITY FOR ENERGY DENSIFICATION | Ultrasonic cavitation |
| PHENOMENON TYPE | Nuclear metamorphosis: Nuclide production |
| MICROPHYSICS INTERPRETATION | In the presence of nuclei of elements present in AISI 304 austenitic steel, under proper stimulation, several isotopes not present before are produced together with nuclear particles, in absence of gamma emissions and in coherence with the Baryon Number Conservation law pertaining to nuclear reactions (which is considered to be valid also in this case). |

***EXPERIMENT TYPE 10.***

***Objective results***

| | |
|---|---|
| MATERIAL | AISI 304 austenitic steel in air |
| STIMULATION TECHNIQUE | Series of pressure cycles at variable rate |
| EXPERIMENTAL EVIDENCES | Alpha particles |
| TECHNIQUES TO DETECT THE EVIDENCES | Electronic ZnS (Ag); Photographic PADC CR39 |

***Inferences***

| | |
|---|---|
| ESTIMATED DEGREE OF DESCRIPTION COMPLETENESS & OF REPRODUCIBILITY LEVEL | Good |
| INTERACTION ENVIRONMENT | AISI 304 austenitic steel |
| INTERACTION AGENT | Elements present in AISI 304 austenitic steel |
| MODALITY FOR ENERGY DENSIFICATION | Time variation of pressure |
| PHENOMENON TYPE | Nuclear emissions |
| MICROPHYSICS INTERPRETATION | In the presence of nuclei of elements present in AISI 304 austenitic steel, under proper stimulation, several isotopes not present before are produced together with nuclear particles, in absence of gamma emissions and in coherence with the Baryon Number Conservation law pertaining to nuclear reactions (which is considered to be valid also in this case). |

*EXPERIMENT TYPE 11.*

*Objective results*

| MATERIAL | Metal powders in deuterium gas |
|---|---|
| STIMULATION TECHNIQUE | Increased pressure of deuterium gas |
| EXPERIMENTAL EVIDENCES | Excess of heat |
| TECHNIQUES TO DETECT THE EVIDENCES | Heat transfer to drive a Sterling motor |

*References*

11.1 *Anomalous difference between reaction energies generated within D2O cell*, Y. Arata and Y. C. Zhang, Japanese Journal of Applied Physics 37, L1274 (1998).

*Inferences*

| ESTIMATED DEGREE OF DESCRIPTION COMPLETENESS & OF REPRODUCIBILITY LEVEL | Good |
|---|---|
| INTERACTION ENVIRONMENT | Deuterium gas |
| INTERACTION AGENT | Metal powders |
| MODALITY FOR ENERGY DENSIFICATION | Compensation of pressure variation by pressure gauge inducing pressure shocks |
| PHENOMENON TYPE | Energy generation |
| MICROPHYSICS INTERPRETATION | In the presence of deuterium inside the bulk of metallic powders, under proper stimulation, excess heat develops and several nuclides not present before are produced together with nuclear particles, in absence of gamma emissions and in coherence with the Baryon Number Conservation law pertaining to nuclear reactions (which is considered to be valid also in this case). |

***EXPERIMENT TYPE 12.***

***Objective results***

| | |
|---|---|
| MATERIAL | Constantan (Cu55 Ni44 Mn alloy) - both nanostructured via electrodeposition and not nanostructured) in the presence of $H_2$ or $D_2$ |
| STIMULATION TECHNIQUE | Electric heating reaching temperature up to 350 °C. |
| EXPERIMENTAL EVIDENCES | Elements generation (C, O, Cl, Ca, and Zn) in Constantan cavities. Excess heat production. |
| TECHNIQUES TO DETECT THE EVIDENCES | SEM equipped with an EDS microprobe Measurements of electric power provided |

***Inferences***

| | |
|---|---|
| ESTIMATED DEGREE OF DESCRIPTION COMPLETENESS & OF REPRODUCIBILITY LEVEL | Very good |
| INTERACTION ENVIRONMENT | Constantan lattice deformed, at the surface of the wire, by the presence of deuterium or hydrogen |
| INTERACTION AGENT | All elements present |
| MODALITY FOR ENERGY DENSIFICATION | Temperature increase |
| PHENOMENON TYPE | Isotope generation Excess heat production |
| MICROPHYSICS INTERPRETATION | In the presence of nickel and other metals, under proper stimulation, several nuclides not present before are produced in absence of emissions of both nuclear particles and gamma rays and in coherence with the Baryon Number Conservation law pertaining to nuclear reactions (which is considered to be valid also in this case). Power is generated as well |

*EXPERIMENT TYPE 13.*

*Objective results*

| MATERIAL | Palladium in the presence of H2 or D2 |
|---|---|
| STIMULATION TECHNIQUE | Temperature increase due to heating via He-Ne laser and excimer laser |
| EXPERIMENTAL EVIDENCES | Elements generation (for instance: C, O, Cl, Ca, and Zn) in cavities with dimensions around tens of micrometers, |
| TECHNIQUES TO DETECT THE EVIDENCES | Coupled plasma mass spectrometry or SEM equipped with an EDS microprobe<br>Measurements of electric power supplied in some cases |

*Inferences*

| ESTIMATED DEGREE OF DESCRIPTION COMPLETENESS & OF REPRODUCIBILITY LEVEL | Very good |
|---|---|
| INTERACTION ENVIRONMENT | Pd lattice deformed by the presence of deuterium or hydrogen |
| INTERACTION AGENT | All metals present |
| MODALITY FOR ENERGY DENSIFICATION | Temperature increase due to laser stimulation |
| PHENOMENON TYPE | Isotope generation |
| MICROPHYSICS INTERPRETATION | In the presence of palladium, under proper stimulation, several nuclides not present before are produced in absence of emissions of both nuclear particles and gamma and in coherence with the Baryon Number Conservation law pertaining to nuclear reactions (which is considered to be valid also in this case). |

***EXPERIMENT TYPE 14.***

***Objective results***

| | |
|---|---|
| MATERIAL | ErD$_3$ and TiD$_2$ |
| STIMULATION TECHNIQUE | Gamma irradiation of deuterium provides photo stripped neutrons |
| EXPERIMENTAL EVIDENCES | Detection of photo-dissociation neutrons and claimed neutrons consistent with DT fusion producing $^3$He and a neutron |
| TECHNIQUES TO DETECT THE EVIDENCES | Neutron detection by EJ-309 liquid scintillator and Stilbene solid state organic scintillator; neutron spectrometry by unfolding methods using HEBROW algorithm |

***Inferences***

| | |
|---|---|
| ESTIMATED DEGREE OF DESCRIPTION COMPLETENESS & OF REPRODUCIBILITY LEVEL | Very Good |
| INTERACTION ENVIRONMENT | Deuterated Erbium and deuterated Ti |
| INTERACTION AGENT | Deuterium and possibly erbium and titanium as well |
| MODALITY FOR ENERGY DENSIFICATION | Pulsed neutrons (generated by photoproduction) produce critical energy density throughout the volume and timing of the neutron bunches. |
| PHENOMENON TYPE | Metamorphosis: Production |
| MICROPHYSICS INTERPRETATION | In the presence of deuterium inside the bulk of metallic samples, under proper stimulation, several nuclides not present before are produced, in absence of gamma emissions, together with neutrons and other nuclear particles, and in coherence with the Baryon number conservation law pertaining to nuclear reactions (which is considered to be valid also in this case). |

**EXPERIMENT TYPE 15.**

**Objective results**

| | |
|---|---|
| MATERIAL | In hydrogen atmosphere, mixture of nickel and Lithium Aluminum hydride |
| STIMULATION TECHNIQUE | Temperature rise with electrical joule effect |
| EXPERIMENTAL EVIDENCES | Excess of heat |
| TECHNIQUES TO DETECT THE EVIDENCES | Thermocouples |

**References**

15.1 *Investigation of the heat generator similar to Rossi reactor,* A. G. Parkhomov, International Journal of Unconventional Science. Reports on Experiments 2015

15.2 *LENR as a manifestation of weak nuclear interactions,* A. G. Parkhomov, International Journal of Unconventional Science. Original research workers 2019
https://drive.google.com/file/d/1UEEBqBpLhiJBBdPbhagQQSWLkmazzJLz/view

**Inferences**

| | |
|---|---|
| ESTIMATED DEGREE OF DESCRIPTION COMPLETENESS & OF REPRODUCIBILITY LEVEL | Sufficient |
| INTERACTION ENVIRONMENT | Hydrogen atmosphere |
| INTERACTION AGENT | Nickel hydride |
| MODALITY FOR ENERGY DENSIFICATION | Compensation, through current regulation of temperature variation in order to obtain heat shocks |
| PHENOMENON TYPE | Thermal energy generation |
| MICROPHYSICS INTERPRETATION | In the presence of hydrogen inside the bulk of Mixture of Nickel and Lithium Aluminum Hydride, under proper stimulation, excess heat develops and several nuclides not present before are produced together with nuclear particles, in absence of gamma emissions and in coherence with the Baryon Number Conservation law pertaining to nuclear reactions (which is considered to be valid also in this case). |

*EXPERIMENT TYPE 16.a*

*Objective results*

| | |
|---|---|
| MATERIAL | Granites with iron in air |
| STIMULATION TECHNIQUE | Impulsive brittle fracture pressure |
| EXPERIMENTAL EVIDENCES | Neutron emission |
| TECHNIQUES TO DETECT THE EVIDENCES | Electronic He$^3$ (long counter) thermodynamic BTI |

*References*

16.1 *Piezonuclear neutrons from fracturing of inert solids*, F. Cardone, A. Carpinteri, G. Lacidogna, Physics Letters A 373 (2009) 4158–4163, 0375-9601 – © 2009 Elsevier B.V. doi:10.1016/j.physleta.2009.09.026

16.2 *Fracto-emissions as seismic precursors*, A. Carpinteri, O. Borla Engineering Fracture Mechanics 177 (2017) 239–250

*Inferences*

| | |
|---|---|
| ESTIMATED DEGREE OF DESCRIPTION COMPLETENESS & OF REPRODUCIBILITY LEVEL | Improvable |
| INTERACTION ENVIRONMENT | Air but not free air; metamorphosis takes place in the air enclosed in cavities contained in granites |
| INTERACTION AGENT | Granites with iron in air |
| MODALITY FOR ENERGY DENSIFICATION | Fracture with cavitation through pressure shocks |
| PHENOMENON TYPE | Nuclear emissions: neutron detection |
| MICROPHYSICS INTERPRETATION | In the presence of iron inside the bulk of granite, under proper stimulation, several nuclides not present before are produced together with nuclear particles, in absence of gamma emissions and in coherence with the Baryon Number Conservation law pertaining to nuclear reactions (which is considered to be valid also in this case). This interpretation, however, is still controversial. |

*EXPERIMENT TYPE 16.b*

*Objective results*

| | |
|---|---|
| MATERIAL | Calcium carbonate without iron in open air |
| STIMULATION TECHNIQUE | Non-impulsive ductile fracture pressure |
| EXPERIMENTAL EVIDENCES | No evidence |
| TECHNIQUES TO DETECT THE EVIDENCES | Electronic He$^3$ (long counter), Thermodynamic BTI |

*References*

16.2 *Fracto-emissions as seismic precursors*, A. Carpinteri, O. Borla, Engineering Fracture Mechanics 177 (2017) 239–250

*Inferences*

| | |
|---|---|
| ESTIMATED DEGREE OF DESCRIPTION COMPLETENESS & OF REPRODUCIBILITY LEVEL | Improvable |
| INTERACTION ENVIRONMENT | Expected: the air enclosed in cavities present in granites |
| INTERACTION AGENT | Expected to be Calcium carbonate (without iron) |
| MODALITY FOR ENERGY DENSIFICATION | Fracture without cavitation (ineffective) |
| PHENOMENON TYPE | Nuclear emissions that were expected but did occur |
| MICROPHYSICS INTERPRETATION | In the absence of proper energy densification no nuclear phenomena occur. |

# EXPERIMENT TYPE 17.

## Objective results

| | |
|---|---|
| MATERIAL | Artificial compounds (silesquioxane[14]) in water with Li and SiC |
| STIMULATION TECHNIQUE | Electrical, photonic and in some cases ultrasonic stimuli |
| EXPERIMENTAL EVIDENCES | Increases in temperature and pressure. Damage to components |
| TECHNIQUES TO DETECT THE EVIDENCES | Calorimetric |

## References

17.1 *A Method to Initiate an LENR Reaction in an Aqueous Solution,* B. Roarty. The 21St International Conference for Condensed Matter Nuclear Science (2018)  https://www.youtube.com/watch?v=G5GHtzI7BGI

## Inferences

| | |
|---|---|
| ESTIMATED DEGREE OF DESCRIPTION COMPLETENESS & OF REPRODUCIBILITY LEVEL | Sufficient |
| INTERACTION ENVIRONMENT | Artificial compounds (silesquioxane[15]) in water |
| INTERACTION AGENT | Lithium, (Li), SiC (Silicon Carbide) |
| MODALITY FOR ENERGY DENSIFICATION | Pulsed time variation of the electric field. Ultrasonic cavitation |
| PHENOMENON TYPE | Production of excess heat |
| MICROPHYSICS INTERPRETATION | In the presence of multiple causes of energy densification an heat excess occurs |

---

[14] Silesquioxanes are inorganic-organic hybrid materials that combine the mechanical, thermal, and chemical stability of ceramics with the solution processing and flexibility of traditional soft materials.